\documentstyle[12pt]{article}
\advance\textheight by \topskip
\begin{document}
\begin{flushright}
SU-ITP-96-17\\
hep-th/9604193\\
\today\\
\end{flushright}
\vspace{1cm}
\begin{center}
\baselineskip=16pt

{\Large\bf    BRANE-ANTI-BRANE ~DEMOCRACY}  \\

\vskip 1.5 cm

{\bf Renata Kallosh${}^{a}$\footnote{E-mail address:
 kallosh@physics.stanford.edu}~
 and ~Arvind Rajaraman${}^{b}$\footnote{E-mail address:
arvindra@dormouse.stanford.edu}}
\vskip 1 cm
${}^{a}$Physics Department, Stanford University, Stanford CA 94305,
USA\\
\vskip 0.4 cm
${}^{b}$Stanford Linear Accelerator Center,\\
    Stanford University, Stanford, California 94309 USA
\\
\vskip 0.7 cm
\end{center}

\vskip 1 cm
\centerline{\bf ABSTRACT}
\begin{quotation}

We suggest a  duality invariant formula for the entropy and temperature of
non-extreme black holes in supersymmetric string theory. The entropy is given
in terms of the duality invariant parameter  of the deviation from extremality   
and 56 SU(8) covariant central charges. It interpolates
between the entropies of  Schwarzschild solution and   extremal solutions with
various amount of unbroken supersymmetries and therefore serves for
classification of black holes in supersymmetric string theories.
We introduce the second auxiliary 56  via  E(7) symmetric constraint.  The
symmetric and antisymmetric combinations of these two
multiplets are related via moduli to the corresponding two
fundamental
representations of E(7): brane and anti-brane ``numbers."  Using  the CPT as
well as C symmetry of the entropy formula and duality one can
explain  the mysterious simplicity of the
non-extreme black hole area formula in terms of  branes and  anti-branes.

 \end{quotation}
\newpage

\baselineskip=15pt

\section{Introduction}

The concept of p-brane democracy was introduced by Townsend \cite{Town}. The
basic idea is that the non-perturbative d=4 string theory treats various
p-branes on equal footing. U-duality \cite{HT}, which is related to
 a discrete
version of E(7) symmetry of N=8 supergravity \cite{CJ},  removes the
distinction between $p=1$ wrapping modes of the string  and other p-branes with
$p>1$ of $d=10$ theory and their wrapping modes. U-duality manifests itself in
particular in  the symmetry of the entropy of extreme black holes under
$E(7;{\bf Z})$ \cite{KK}. As a result of  1/8 of unbroken supersymmetry one can
prove \cite{FK2} that the entropy depends only on  the conserved charges
which transform as one 56-dimensional
fundamental multiplet of E(7).
 The entropy is
therefore
given by the unique quartic invariant of E(7) constructed from one fundamental
multiplet of charges. The independence of  the   moduli at fixed values of
conserved charges follows directly from unbroken supersymmetry. An analogous
picture was found in $d=5$ where the corresponding entropy is given  by a
triple invariant of $E(6)$ constructed from one 27-dimensional fundamental
representation of E(6) \cite{FK1,FK2}. It is worth mentioning that in each case
the unique expression for the entropy is protected by the supersymmetric
non-renormalization theorem from quantum corrections.

In view of these facts it looked rather puzzling that  the entropy formulas for
non-extreme black holes in $d=5$ \cite{HMS} and in $d=4$ \cite{HLM}   have been
recently suggested in a form of triple and quartic invariants respectively
based in each case on  two fundamental representations of E(6) (E(7)) in
$d=5$ $(d=4)$.  The first 27 (56) in d=5 (d=4)
corresponds to the number of solitons and the second one to the number of
anti-solitons.

The purpose of this letter is first to obtain the duality invariant entropy and
temperature formulas for  non-extreme black holes of N=8 supergravity in d=4.
We will present a simple formula for the product of the entropy and temperature
$ST$ in terms of the eigenvalues of the supersymmetry charges.
The formula for the entropy will involve only the mass of the black hole and
the central charge
matrix and will generalize the previously known formulas of this kind for N=4
supergravity.
There is nothing puzzling about this formula since there is only one $N\times
N$ antisymmetric complex central charge matrix $Z_{AB}$ with $ A,B =1, \dots ,
N $ in the theory and no obvious room for branes and anti-branes. The central
charge is invariant under E(7) and  transforms covariantly under the global
$SU(8)$ symmetry.

Starting from this formula we will introduce the second  $SU(8)$ covariant
charge $Y_{AB}$ as a function of the parameter of the deviation of the theory
from extremality $r_0$ and  of the central charge  $Z_{AB} $.   The total
procedure is very much in spirit of Dirac's treatment of constrained systems:
originally one has more variables then necessary to describe the physical
states. The symmetries of the theory have enough room to be manifest
when all of these variables are present. The variables are constrained,
however. If one would like to describe the system only in terms of
unconstrained variables one has to solve the constraints and break some of the
symmetries.

Our auxiliary variable $Y_{AB} $ will serve the following purpose:
whereas the central charge  $Z_{AB}  $ will be represented as an antisymmetric
function of branes and anti-branes, which changes its sign when
branes are changed into anti-branes, the  auxiliary variable $Y_{AB}$
will be constructed as a symmetric function of  branes and anti-branes.

Starting with our two 56 of SU(8)
which have a very clear interpretation via an E(7) symmetric constraint between
Z's and Y's,
we will then
rewrite our entropy formula in terms of  a symmetric function of 56
``brane-numbers"
and 56 ``anti-brane-numbers"  as in \cite{HMS,HLM} using the
quartic invariant of E(7) applied to two multiplets.
 In
this way   we will confirm the entropy formulas suggested in \cite{HMS,HLM} and
explain
the  origin  of the two multiplets in a covariant manner.

Moreover, our way of derivation of the formulas of the
non-extreme black holes will suggest the new interpretation which will go under
the name ``brane-anti-brane democracy" and will state not only the equal rights
of various p-branes but also the equal rights of states with positive and
negative charges. This can be understood as follows. For the non-extreme
configurations the integer  charges
$(p^I, q_I), \; I=1, \dots , 28$,
  will be given by the antisymmetric combinations of branes and anti-branes
\begin{equation}
p^I =  m^I  - \bar m^I \ ,  \qquad  q_I=   n_I - \bar n_I \ .
\end{equation}
This means that the charges can be positive or  negative, depending on whether
$n>\bar n$ and $m> \bar m$ or vice versa. The non-extreme entropy will depend
on $(m^I, n_I)$  and on $(\bar m^I, \bar n_I)$ in some symmetric way, which
reflects the degeneracy of the non-extreme black holes entropy on the sign of
charges.

For the extreme solutions again the charges can be positive  or negative in
each of the 28 groups, the entropy
being degenerate in this signs.   Note that  the  brane $(m^I, n_I)$
 and anti-brane numbers  $(\bar m^I, \bar n_I)$
can  only be non-negative numbers in the extreme limit.
The charges are given by the number of  branes
\begin{equation}
(p^I )_{\rm extr} = (|p^I |)_{\rm extr} = m^I\ ,  \qquad  (q_I)_{\rm extr} =
(|q_I
 |)_{\rm extr} = n_I\ ,
\end{equation}
for those values of $I$ for which the charges are positive and by the number of
anti-branes
\begin{equation}
(p^I )_{\rm extr} = (- |p^I |)_{\rm extr} = - \bar m^I\ ,  \qquad  (q_I)_{\rm
extr} = (- |q_I  |)_{\rm extr} = - \bar n_I \ ,
\end{equation}
for those values of $I$ for which the charges are negative. Therefore the
entropy for all solutions with all possible values of charges in all gauge
groups   in extreme limit will depend only on absolute values of all charges
$(|p^I|, |q_I|)$.
Since we will consider even non-extreme black holes in the context of
supersymmetric theories, the CPT as well as C transformation will be defined in
terms of the transformations of the supersymmetry generators. The CPT
transformation on supersymmetry charges acts as follows: $Q  \rightarrow i
Q^*$.
This results in $P_\mu \rightarrow
P_\mu$ and $Z \rightarrow  - Z^*$. This is realized  in terms of charges as
$p\rightarrow p$ and
$q \rightarrow -q$. Therefore CPT on branes and anti-branes act as follows:
\begin{equation}
n \Longleftrightarrow  \bar n \ ,   \qquad m \Longleftrightarrow   m\ ,  \qquad
\bar m
\Longleftrightarrow  \bar m \ .
\end{equation}
The charge conjugation C acts on supersymmetry generators as follows:  $Q
\rightarrow i Q $
and  $Q^*   \rightarrow  - i Q^*$, which results in  $P_\mu \rightarrow
 P_\mu$,\,  $Z \rightarrow  - Z $,\, $Z^*  \rightarrow  - Z^*$.  This leads to
\begin{equation}
n \Longleftrightarrow  \bar n \ ,  \qquad m \Longleftrightarrow  \bar m  \ .
\end{equation}

\section{Supersymmetry algebra, temperature and entropy of non-extreme black
holes}

There exists a simple relation \cite{US} between the product of temperature T
and entropy S of black holes in supersymmetric theories and the  parameter of
the deviation of the theory from extremality $r_0 $:
\begin{equation}
2 \pi S\, T = r_0 \equiv {r_+ - r_- \over 2} \ .
\end{equation}
This parameter  $r_0$ defines the distance between the event horizon $r_+$ and
the inner horizon $r_-$ of the non-extreme black holes.

We will  derive the universal  formula  for $2 \pi S\, T = r_0$ in terms of
supersymmetry charges by using  the symplectic form of the supersymmetry
algebra  \cite{FSZ}. The d=4 N-extended supersymmetry algebra   is most
conveniently described for massive states at rest  in terms of  $2N$-component
spinors.  In doublet form they are given by
\begin{equation}
 Q_{\alpha}^{ a} = \left (\matrix{
Q_{\alpha  A} \cr
\cr
 Q^{* \alpha   A}\cr
}\right ), \quad \matrix{
Q_{\alpha}^{ a} = Q_{\alpha A} \  \  \ \mbox{for} \ \ \
a=1,\dots ,  N , \qquad \alpha,\beta =1,2 \ ,
\cr
\cr
Q_{\alpha}^{ a}= Q^{* \alpha   A} =\epsilon^ {\alpha \beta} Q^{* A}_
\beta \  \  \ \mbox{for} \ \ \
a= N+1,\dots , 2N \ .
\cr
}
\end{equation}
These spinors satisfy a symplectic reality condition
$
Q_{\alpha}^{*a} = \epsilon ^{\alpha\beta} \Omega_{ab} Q_{\beta}^{ b}
$
with
\begin{equation}
\epsilon^{\alpha \beta } = - \epsilon_{\alpha \beta } = (i \sigma_2)_{\alpha
\beta }   , \qquad \Omega^{ab} = - \Omega_{ab}  = \pmatrix{
0&  {\rm I}\cr
-   {\rm I} & 0 \cr } \  , \end{equation}
and the supersymmetry algebra in a symplectic form is
\begin{eqnarray}
\{  Q_{\alpha}^a ,  Q_{\beta }^b \} = \epsilon _{\alpha \beta} \pmatrix{
Z  & M {\rm I}\cr
\cr
- M  {\rm I} & Z^{*} \cr } \equiv  \epsilon _{\alpha \beta} \; \Lambda ^{ab}
\label{Z} \ .
\end{eqnarray}
The $2N\times 2N$ matrix $\Lambda ^{ab}$ is written in terms of $N \times N$
numerical antisymmetric complex matrix $Z_{AB}, , Z^{*AB}$ and the mass. The
numbers $Z_{AB}$ are the eigenvalues of the central charge operators in a given
supermultiplet.
Any complex antisymmetric matrix $Z_{AB}$ can be brought to the normal form
using some $U(N)$ rotation \cite{Z}.  For example,  for $N=4$ and for $N=8$
respectively
we get
\begin{equation}
\tilde Z_{AB} = i \sigma_2 \left (\matrix{
z_1 & 0  \cr
0 & z_2  \cr
}\right )  , \qquad
\tilde Z_{AB } = i \sigma_2
\left (\matrix{
z_1 & 0 & 0 & 0 \cr
0 & z_2 & 0 & 0 \cr
0 & 0 & z_3 & 0 \cr
0 & 0 & 0 & z_4 \cr
}\right )  ,
\end{equation}
where $z_i$ are  non-negative real numbers, $i= 1, \dots N/2$. We used the
following notation: $\tilde {Z}_{12} = -\tilde {Z} _{21} = z_1,  \dots ,
\tilde {Z}_{78} = -\tilde {Z} _{87} = z_4$.

In terms of the mass of the black hole and the central charge
the product of the entropy and temperature for N=4 case can be written as
\begin{equation}
(2 \pi S\, T)^2    = r_0^2\equiv {(M^2- |z_1|^2) (M^2- |z_2|^2)\over M^2} \geq
0 \ ,
\label{N=4}\end{equation}
and for N=8 case as
\begin{equation}
(2 \pi S\, T)^2 = r_0^2\equiv {(M^2- |z_1|^2) (M^2- |z_2|^2) (M^2- |z_3|^2)
(M^2- |z_4|^2)\over M^6} \geq 0 \ .
\label{N=8}\end{equation}

 The product of
temperature and entropy then has a simple representations in terms of the
eigenvalues of the supersymmetry generators.

\begin{equation}
(2 \pi S\, T)^2  = r_0^2   = {det^{1/2}  \{  Q_{\alpha  }^c ,  Q_ {\gamma}^ b
\} \over [{1\over 4N}
\Omega_{ac} \epsilon^ { \beta \gamma}
\{  Q_{\alpha  }^c ,  Q_ {\gamma}^ b  \}]^{N-2} } \ .
\label{ST} \end{equation}

This formula is completely symmetric in terms of  32 supersymmetry generators
 of N=1, d=11 supersymmetry in case we study N=8, d=4 theory and
symmetric in  16 supersymmetry generators of N=1, d=10 theory when we study
N=4, d=4 theory.

As long as $2 \pi S\, T = r_0  \neq 0$ the black holes are  non-extreme. The
extreme ones   have $r_0=0$, which means that some supersymmetry charges have
to have zero eigenvalues. Those with the non-vanishing area of the horizon and
entropy have vanishing
temperature.
\begin{equation}
r_0 =0 \ , \qquad T=0 \ , \qquad S\neq 0 \ .
\end{equation}
The extreme ones with the vanishing area of the horizon $S=A/4\pi = 0$ usually
do not have a well defined temperature since the horizon is singular.

The analysis performed in this section shows that one can describe the geometry
 starting with the simplest solutions related to the normal form of the
central charge matrix  $Z_{AB}$ and derive the formulas of the type
(\ref{N=4}), (\ref{N=8}).  The result can then be generalized to the form in
which it is no longer necessary to assume  that the central charge matrix is
diagonal.

\section{Thermodynamics of black holes in N=2,4,8 supergravity}

In this section we will use the known results for the non-extreme black holes
in N=4 supergravity \cite{US} and represent them in the form suitable for
generalization to N=8 theory. The entropy given by
1/4 of the area was found in \cite{US} to be equal to
\begin{equation}
S =\pi   ( r_0 + M +\Sigma  ) ( r_0 +M -  \Sigma ) \ ,
\label{entr1}\end{equation}
where  $\Sigma$ is  the dilaton  charge of the black hole.
 Using the fact that
$$\Sigma =-  {z_1 z_2 \over M}$$
 one can show that the entropy
 is equal to
\begin{equation}
S_{N=4} = \pi \left ( r_0 + \sqrt {r_0^2 + (z_1 + z_2 )^2} \right ) \left ( r_0
+
\sqrt {r_0^2 + (z_1 - z_2 )^2} \right ) \ .
\label{entr2}\end{equation}
It is easy to get the entropy formula in pure  the N=2 supergravity by setting
$z_2=0$
\begin{equation}
S_{N=2} = \pi \left ( r_0 + \sqrt {r_0^2 + (z_1  )^2} \right )^2 \ .
\label{entr3}\end{equation}
The generalization of N=4 formula to the N=8 theory is  straightforward
\begin{eqnarray}
S_{N=8}  =  \pi \left ( r_0 + \sqrt {r_0^2 + (z_1 + z_2 +z_3 +z_4)^2} \right
)^{1/2} \left ( r_0 +
\sqrt {r_0^2 + (z_1 + z_2 -z_3 -z_4)^2} \right )^{1/2}&&\nonumber\\
  \left ( r_0 + \sqrt {r_0^2 + (z_1 - z_2 +z_3 -z_4)^2} \right )^{1/2} \left (
r_0 + \sqrt {r_0^2 + (z_1 - z_2 -z_3 +z_4)^2} \right )^{1/2}&.&
\label{entr4}\end{eqnarray}
This   expression  agrees with  the   non-extreme black hole
solution with four-charges  \cite{CY} upon
 identification of central charges of N=8 theory performed in \cite{KK}.
There is one more important criterion for the validity of this formula: at the
extreme limit $r_0=0$
the mass of the solution becomes equal to the largest of the eigenvalues, for
example, $z_1 \equiv Z$ ,   which depends on the conserved charges and on
moduli:
\begin{equation}
M_{\rm extr} = |Z \left ((p,q) ,\phi_{\infty} \right )| \ .
\end{equation}
According to universality of the supersymmetric attractors
\cite{FK2}, near the horizon all central charges besides the largest one $|Z|$
 have to
vanish,  i.e.
$z_2=z_3=z_4=0$,  and any of these expressions for N=2,4,8 has to reduce to
\begin{equation}
S_{\rm extr} = \pi |Z \left (( p, q) , \phi_{\rm h}  [(p,q)]\right )| \ ,
\end{equation}
so that the entropy depends only on conserved charges (56 in N=8,\, 12 in
N=4,\, 2
in N=2).
The central charge $Z_{AB}$  enters the local supersymmetry transformations
rules, since it is a charge of the graviphoton.

To switch from the SU(8) basis to the SO(8) basis in the general case of N=8
theory when
neither $Z_{AB}$ nor $\zeta_{ij} $ are in the normal form, we will use
\begin{equation}
\zeta_{ij}= {1\over 2\sqrt 2} Z_{AB} (\gamma^{ij})_A{}^B
\end{equation}
with the inverse relation
\begin{equation}
Z_{AB}= {1\over 4\sqrt 2} \zeta_{ij} (\gamma^{ij})_A{}^B \ ,
\end{equation}
 where the matrices $ (\gamma^{ij})_A{}^B$, $i,j = 1, \dots , 8 $, form  the
algebra of $SO(8)$.
In the normal frame they corresponds to linear combinations
\begin{eqnarray}
 \zeta_1\equiv  \zeta_{12} =
{1\over \sqrt 2}(z_{1} +z_{2}+z_3 +z_4)\ , \qquad  \zeta_2\equiv \zeta_{34} =
{1\over \sqrt 2}(z_{1} +z_{2} -z_3 -z_4)\ ,\nonumber\\
\zeta_3\equiv  \zeta_{56} =
{1\over \sqrt 2}(z_{1} -z_{2}+z_3 -z_4)\ , \qquad  \zeta_4\equiv \zeta_{78} =
{1\over \sqrt 2}(z_{1} -z_{2} +z_3 -z_4) \ .
\end{eqnarray}
 The entropy formula can now be rewritten as follows:
\begin{equation}
S= \pi (\det )^{1/4}  \left\{ r_0 \delta_{ik} + \sqrt {(r_0)^2 \delta_{ik}  - 2
\zeta_{ij} \zeta^* _{jk}} \right\} \ .
\label{entr5}\end{equation}
Here we can also use the diagonal basis, and consider the case where only
$\zeta_{1}= \zeta_{2}$ and  $\zeta_{3}= \zeta_{4}$ are not vanishing (they
still can be complex) and we get
\begin{equation}
S=\prod_{i=1}^{4}  \left\{ r_0  + \sqrt {(r_0)^2   + 2
\zeta_{i} \zeta^* _{i}} \right\}^{1/2} \ .
\end{equation}
This formula describes in particular the entropy of the non-extreme $U(1)\times
U(1)$ axion-dilaton black holes in which both gauge groups have electric as
well as magnetic charges \cite{US}.

Now we can introduce the auxiliary $SU(8)$ multiplet   $(Y_{AB}, Y^{*AB})$,
which is a function of   $r_0$ and $(Z_{AB}, Z^{*AB})$. First, let us again
switch to the $SO(8)$ basis
\begin{equation}
Y_{AB}= {1\over 4\sqrt 2} \Upsilon_{ij} (\gamma^{ij})_A{}^B \ ,\qquad
\Upsilon_{ij}= {1\over 2\sqrt 2} Y_{AB} (\gamma^{ij})_A{}^B \ .
\end{equation}
The definition of our new multiplet $Y$ is the following:
\begin{equation}
(r_0)^2 \delta_{ik}  - 2 \zeta_{ij} \zeta^* _{jk} \equiv - 2\Upsilon _{ij}
\Upsilon^* _{jk} \ .
\label{constr}\end{equation}
Thus we have a non-linear relation (\ref{constr})  between the parameter of
deviation of extremality, central charges of the theory $(Z_{AB}, Z^{*AB})$ and
new auxiliary  $SU(8)$ multiplet   $(Y_{AB}, Y^{*AB})$.
The expression for the entropy which was given only in terms of $r_0$ and
central charges can be rewritten  using the auxiliary multiplet $\Upsilon$ as
\begin{equation}
S= \pi (\det )^{1/4}  \left\{ r_0 \delta_{ik} + \sqrt  {- 2 \Upsilon_{ij}
\Upsilon^* _{jk}} \right\} \ .
\label{entr6}\end{equation}

It will be useful from now to use the  basis with  28 real electric and  28
real magnetic charges
\begin{eqnarray}
\zeta_{ij} &= &Q_{ij} + i P^{ij} \ ,\qquad  \Upsilon_{ij} = Q'_{ij} + i P'^{ij}
\ , \\
\nonumber\\
\zeta^{* ij} &=& Q_{ij} - i P^{ij} \ , \qquad  \Upsilon^{*ij} =Q'_{ij} - i
P'^{ij} \ .
\end{eqnarray}
The charges $ A\equiv (P^{ij} ,  Q_{ij}) $ are related to the conserved charges
$a \equiv (p^{ij} ,  q_{ij}) $ via moduli  $A= Va$.  For our auxiliary
multiplet
 $ A'\equiv (P'^{ij} ,  Q'_{ij}) $ we will assume the same relation $A'= Va'$
with
$a' \equiv (p'^{ij} ,  q'_{ij}) $, it is consistent with the constraint
(\ref{constr}).
One more step is to define the symmetric and antisymmetric combinations of
the original and the auxiliary  multiplets:
\begin{eqnarray}
P'^{ij} +  P^{ij} &= &M^{ij} \ , \qquad Q'_{ij} + Q_{ij} =N_{ij} \ ,\\
P'^{ij} -  P^{ij} &=& \bar M^{ij} \ , \qquad Q'_{ij} - Q_{ij} =\bar N_{ij} \ .
\end{eqnarray}
Our  constraint (\ref{constr}) simplifies to
\begin{equation}
(r_0)^2 \delta_{ik}  = -  2 ( M^{ij} \bar M^ {jk} + N _{ij}
\bar N_{jk}) \ .
\label{constr3}\end{equation}
One can exclude $r_0$ completely, using the constraint
(\ref{constr3}),
and rewrite the entropy formula entirely in terms of these two 56s of $SU(8)$
$(M,N) $ and $(\bar M, \bar N)$
as
follows:
\begin{equation}
S= 2 \pi (\det )^{1/4}  \left\{ 2  \sqrt { M^{ij}  \bar M^{jk} + N_{ij} \bar
N_{jk} }  + \sqrt  { (M +\bar M)^{ij}  (M +\bar M)^{jk} + (N +\bar N)_{ij} (N
+\bar N_{jk}) } \right\} \ .
\label{entrMN}\end{equation}

To rewrite this formula in terms of 56 branes and 56 anti-branes we have to
review some elements of N=8 theory \cite{CJ}.
We consider the  theory in the symmetric gauge with fixed local $SU(8)$
symmetry. In this gauge the  scalars are taken to be the coordinates of the
coset space ${E_7 \over SU(8)}$. The matrix $V$ describing the scalars forms an
element of E(7) before the local gauge  fixing and has 133 entries, but  when
unitarity constraint
$V=V^\dagger$  is imposed,   it depends only on 35 complex scalars. In this
gauge the hidden symmetry acts on the  charges as well as on moduli as follows
\begin{equation}
a= \left (\matrix{
p\cr
q\cr
}\right )\ , \qquad a \longrightarrow E a\ , \qquad V \longrightarrow  h V
E^{-1} \ ,
\qquad
A=Va  = \left (\matrix{
P\cr
Q\cr
}\right ) \longrightarrow   h A \ .
\end{equation}
Here $E$ is an element of $E(7)$ and $h$ is the global $SU(8)$ which is the
residual symmetry after gauge fixing, which preserves the gauge. It follows
that the global symmetry of  $A=V a$ is the $SU(8)$ symmetry.  The same
transformation rules apply to the
the auxiliary multiplet of E(7) ,  $a'$ and the corresponding $SU(8)$ partner
of it with E(7) blind, which transforms only under $SU(8)$,   $A'$.
\begin{equation}
a'= \left (\matrix{
p'^I\cr
q'_I\cr
}\right )\ , \qquad a \longrightarrow E a'\ , \qquad V \longrightarrow  h V
E^{-1}\ ,
\qquad
A'=Va' = \left (\matrix{
P'\cr
Q'\cr
}\right ) \longrightarrow   h A' \  .
\end{equation}
As already explained in the Introduction, we will build a symmetric and
antisymmetric combination of our two E(7) multiplets and call them branes and
anti-branes,
respectively.
\begin{equation}
a= \left (\matrix{
 p^{ij}\cr
q_{ij}\cr
}\right )= \left (\matrix{
(m -\bar m)^{ij} \cr
(n -\bar n)_{ij}\cr
}\right )\ , \qquad    a'= \left (\matrix{
p'^{ij}\cr
q'_{ij}\cr
}\right )= \left (\matrix{
(m +\bar m)^{ij} \cr
(n +\bar n)_{ij}\cr
}\right )\ .
\end{equation}
The symmetric and antisymmetric $SU(8)$ partners of these two E(7) multiplets
are
\begin{equation}
A'+A = \left (\matrix{
M^{ij}\cr
N'_{ij}\cr
}\right )\ , \qquad    A'-A = \left (\matrix{
\bar M^{ij}\cr
\bar N'_{ij}\cr
}\right )\ .
\end{equation}
The moduli matrix $V=V^\dagger $ in the symmetric gauge is a function of the
matrix   $y_{ij, kl}$ which defines
 the  inhomogeneous coordinates of
${E_7
\over SU(8)}$.
 E(7)  acts on the 70 coordinates $y_{ij, kl}$  by fractional transformation
\begin{equation}
y' = {B+yD \over A+yC} \ .
\label{frac}\end{equation}
and
 $A, B, C, D $ are 28 by 28  constant matrices defined in \cite{CJ}.

Having introduced all these objects, which in addition to charges contain an
auxiliary multiplet $a'$,  we will proceed  from the other side and use the
entropy formulas  (\ref{entr5}), (\ref{entr6}).
The question is whether our  entropy (\ref{entrMN})
$$S \left \{ (M,N) , (\bar M, \bar N)\right\} $$
can be proved to depend only on
$(m,n) $ and $(\bar m, \bar n)$ and not on moduli $V$. Equivalently one may try
to show that the 70 moduli are functions of $(m,n) $ and $(\bar m, \bar n)$. By
using only the E(7) duality symmetry of the theory or the $SU(8)$ subgroup of
it  we cannot prove that the entropy does not depend on moduli. However,
fortunately, the symmetry of the entropy is larger: the original formula
(\ref{entr5}) for the entropy in terms of central charges in addition to an
$SU(8)$ symmetry of $\zeta_{ij}, \zeta^{*ij}$
has a CPT symmetry  under $\zeta_{ij} \longrightarrow - \zeta^{*ij}$,  C
symmetry under
$\zeta_{ij} \longrightarrow - \zeta_{ij}$ as well as a U(1) symmetry
$\zeta_{ij} \longrightarrow - e^{i\alpha} \zeta^{ij}$.
Therefore we can use the  global $U(8)$ symmetry of the entropy.

The entropy for the most general solution can be therefore  reduced  to the
expression which it has for the simplest solution in the normal frame.
\begin{equation}
S= 2 \pi (\det )^{1/4}  \left\{ r_0 \delta_{ik} + \sqrt  {- 2 \Upsilon_{ij}
\Upsilon^* _{jk}} \right\} =  2 \pi (\det )^{1/4}  \left\{ r_0 \delta_{ik} +
\sqrt  {- 2 \tilde \Upsilon_{ij}
\tilde \Upsilon^* _{jk}} \right\} \ .
\end{equation}
This can be further presented as
\begin{equation}
S= 2 \pi (\det )^{1/4}  \left\{ 2  \sqrt {  N_{ij} \bar N_{jk} }  + \sqrt  {  +
(N +\bar N)_{ij} (N +\bar N)_{jk} } \right\} \ ,
\end{equation}
which is equal to
\begin{equation}
S=2\pi(\sqrt{N_{12}}+\sqrt{\bar{N}_{12}})(\sqrt{N_{34}}+\sqrt{\bar{N}_{34}})
(\sqrt{N_{56}}+\sqrt{\bar{N}_{56}})(\sqrt{N_{78}}+\sqrt{\bar{N}_{78}}) \ .
\end{equation}
 It has been  found in \cite{HLM} that 3 moduli of the normal frame solution
are functions of 8
branes and anti-branes.
Therefore it was possible to  present an entropy  in a form where it depends
only on $\tilde n, \tilde {\bar n}$ which are the normal form representative of
the E(7) multiplets $(m, n)$, $(\bar m,\bar n)$.
\begin{equation}
S=2\pi(\sqrt{n_{12}}+\sqrt{\bar{n}_{12}})(\sqrt{n_{34}}+\sqrt{\bar{n}_{34}})
(\sqrt{n_{56}}+\sqrt{\bar{n}_{56}})(\sqrt{n_{78}}+\sqrt{\bar{n}_{78}}) \ .
\label{HLM}\end{equation}
It was suggested in \cite{HLM}, following the related observation in d=5
theory in \cite{HMS}, to consider the  entropy formula, generalizing the
normal frame solution to an E(7) invariant form.
\begin{equation}
S=2\pi\sum_{i,j,k,l} \sqrt{T_{\hat A \hat B \hat C \hat D }f_i^{\hat A}
f_j^{\hat B} f_k^{\hat C} f_l^{\hat D}}\ , \qquad i,j,k,l=1,2\ , \qquad  \hat A
, \hat
B,  \hat C,  \hat D = 1,\dots , 56 \ ,
\label{cartan2}\end{equation}
where $T_{\hat A \hat B \hat C \hat D}$ is the quartic invariant considered in
\cite{KK}, and
$f_1^{\hat A }= (m^{ij}, n_{ij})$,  $f_2^{\hat A }= (\bar m^{ij}, \bar
n_{ij})$.
The mysterious part of this formula was the appearance of two  full E(7)
multiplets, and the interpretation of this multiplets  since the
classical solution depends only  28 vector fields.
Now we are in a position to prove  this formula for the entropy of non-extreme
black holes and explain the origin of the second multiplet.

We start  with our U(8) symmetric formula (\ref{entrMN}) which depends on two
SU(8) multiplets and is invariant under E(7). We could rewrite this
formula as a function of the moduli and the ``brane-numbers" and
``anti-brane numbers"  ($m,n,\bar{m},\bar{n}$) i.e.
\begin{equation}
S \left \{ (M,N) , (\bar M, \bar N)\right\}=\hat S \left \{ (m,n) , (\bar
m,\bar n), y_{ij, kl} \right\} \ .
\end{equation}
However,  we see from equation (\ref{HLM}) that the entropy in
the normal frame can be written as a function of the ``brane-numbers"
and ``anti-brane-numbers" alone. This shows that in the normal frame, all the
moduli appearing in the function $\hat S$ are determined as functions of
($n,\bar{n}$).
\begin{equation}
 S_{\rm normal} \left \{ (0, \tilde n) , (0, \tilde {\bar n}) \right\}  = \hat
S_{\rm normal} \left \{ (0, \tilde n) , (0, \tilde {\bar n}) , \tilde {y} _{ij,
kl}( \tilde n,  \tilde {\bar n})\right\} \ .
\label{proof2}\end{equation}
The E(7) symmetric generalization of the left hand side of this equation due to
the uniqueness of the quartic invariant \footnote{One could also have a
symplectic quadratic invariant of E(7) but
it would not have a  correct supersymmetric limit and will not match the normal
frame solution} is given by the formula (\ref{cartan2}). For this to be
consistent with the right hand side of eq. (\ref{proof2})  moduli in this
formula have to transform under E(7) via it dependence on $\tilde n,  \tilde
{\bar n}$. Therefore the transformed value of the moduli according to eq.
(\ref{frac}) will provide the generic expression for the 70 moduli  as
functions of all branes and anti-branes
$y_{ij, kl}\left( (m,n) , (\bar m, \bar n)\right)$.

Thus we conclude that entropy of the the non-extreme black holes is duality
invariant and has a nice symmetric C-invariant form given in eq.
(\ref{cartan2}) as conjectured in \cite{HLM}. Under C-conjugation
\begin{equation}
f_1 \Longleftrightarrow f_2 \ .
\end{equation}
Thus the entropy can indeed be written entirely in terms of branes and
anti-branes, which however are not independent but satisfy an  E(7) symmetric
constraint (\ref{constr3}), which allows to express the moduli in terms of
branes and anti-branes:
\begin{equation}
(r_0)^2 \delta_{ik}  = -  2  ( M^{ij} \bar M^ {jk} + N _{ij}
\bar N_{jk}) = -2 \left(  (\bar a^\dagger  V^\dagger )^{ij}  (Va)^{jk}  + (\bar
a^\dagger  V^\dagger )_{ij}  (Va)_{jk} \right) \ .
\label{constr4}\end{equation}
In the normal frame  this set of constraints
becomes
\begin{eqnarray}
r_0^2 = 2 N _{12}
\bar N_{12} = 2 N _{34}
\bar N_{34} =2 N _{56} \bar N_{56}= 2 N _{78} \bar N_{78} \ .
\end{eqnarray}
Together with diagonal form of relations  $A =Va$, $A' = Va'$
\begin{equation}
N _{12}= V_{12}{}^{12} n_{12}\ , \qquad N _{34}= V_{34}{}^{34} n_{34}\ , \qquad
N
_{56}= V_{56}{}^{56} n_{56}\ , \qquad N _{78}= V_{78}{} ^{78} n_{78} \ ,
\end{equation}
\begin{equation}
\bar N _{12}=
V_{12}{}^{12}
 \bar n_{12}\ , \qquad \bar N _{34}= V_{34}{}^{34} \bar n_{34}\ , \qquad \bar N
_{56}= V_{56}{}^{56} \bar n_{56}\ , \qquad \bar N _{78}= V_{78}{} ^{78} \bar
n_{78} \ ,
\end{equation}
this leads to
\begin{equation}
(V_1)^2 n_1 \bar n _1 = (V_2)^2 n_2 \bar n _2= (V_3)^2 n_3 \bar n_3 =(V_4)^2
n_4 \bar n_4\ ,
\end{equation}
where as before we have simplified the notation, e.g. $n_{12} \equiv n_1,
V_{12}{}^{12} \equiv V_1 $.
Thus we have derived from our U(8) covariant constraint the fact established in
\cite{HLM} that three values of moduli,
\begin{equation}
\left ( {V_2 \over V_1}\right )^2 = {n_1 \bar n _1 \over n_2 \bar n _2}\ ,
\qquad
\left ( {V_3 \over V_2}\right )^2 = {n_2 \bar n _2 \over n_3 \bar n _3}\ ,
\qquad
\left ( {V_4 \over V_3}\right )^2 = {n_3 \bar n _3 \over n_4 \bar n _4} \ ,
\end{equation}
are the functions of 4 branes and 4 anti-branes.  This seemed earlier to be a
mysterious property of the solution. We have established this property by
introducing into the theory an auxiliary SU(8) multiplet constrained to the
original one in E(7) symmetric way. This has allowed to realize the degeneracy
of the entropy formula in the
signs of the charges in each of 28 gauge groups in a manifest way. Thus the
entropy formula (\ref{cartan2}) given in terms of two E(7) multiplets by
Cartan's quartic invariant, when the 2 multiplets are constrained as in eq.
(\ref{constr4}),  is the entropy formula for non-extreme black holes in N=8
supergravity.

\vskip 1 cm

In conclusion, we have found   entropy formulas  for N=8 supergravity black
holes in terms of central charges (\ref{entr4}), (\ref{entr5}) which
interpolate between Schwarzschild solution with all central charges vanishing
$Z_{AB}=0$, Reissner-Nordstr\"om  solutions with one non-vanishing skew
eigenvalue of the central charge and various other non-extreme solutions.
 It simultaneously includes  those with
1/8, 1/4 and 1/2 of unbroken supersymmetry, depending on how many of the
supersymmetric positivity bounds are saturated, i.e. whether $M=z_1>z_2, z_3,
z_4$,
or  $M=z_1=z_2> z_3, z_4$,  or   $M=z_1=z_2= z_3= z_4$.

This formula, together with the expression for the product of the temperature
and entropy  in terms of supersymmetry charges
(\ref{ST}), presents  the classification of extreme as well as non-extreme
black holes in supersymmetric theories. This kind of classification was
suggested before in the limited context of pure N=4 supergravity, when only
$z_1, z_2$ where available
\cite{US}.

We have also rederived this entropy formula in terms of conserved charges of
the theory and moduli and, finally,  in terms of branes and anti-branes.
We gave an explanation
for the appearance of branes and anti-branes in the description of these
black holes as a set of constrained multiplets which enable us to realize
the full symmetries of the theory.

\section{Acknowledgements}
 Stimulating discussions with B. Kol,  A. Linde and  L.
Susskind are gratefully acknowledged.
The work of R.K. is supported by the  NSF grant
PHY-9219345. The work of A.R is supported in part by the Department of Energy
under contract no. DE-AC03-76SF00515.

\end{document}